\documentclass[prl,aps,preprint,noshowpacs,superscriptaddress]{revtex4}
\usepackage{dcolumn}
\usepackage{bm}
\usepackage{amsmath,amssymb}
\usepackage{graphicx}

\usepackage{color}

\begin{document}

\title{Comment on ``How to Observe Coherent Electron Dynamics Directly'' 
[H. J. Suominen and A. Kirrander, Phys. Rev. Lett. {\bf 112}, 043002 (2014)]}
\author{Robin Santra}
\affiliation{Center for Free-Electron Laser Science, DESY, Notkestra{\ss}e 85, 22607 Hamburg, Germany}
\affiliation{Department of Physics, University of Hamburg, Jungiusstra{\ss}e 9, 20355 Hamburg, Germany}
\author{Gopal Dixit}
\affiliation{Max Born Institute, Max-Born-Stra{\ss}e 2 A, 12489 Berlin, Germany}
\author{Jan Malte Slowik}
\affiliation{Center for Free-Electron Laser Science, DESY, Notkestra{\ss}e 85, 22607 Hamburg, Germany}
\affiliation{Department of Physics, University of Hamburg, Jungiusstra{\ss}e 9, 20355 Hamburg, Germany}
\date{\today}
\begin{abstract}
This is a comment on ``How to Observe Coherent Electron Dynamics Directly'' 
[H. J. Suominen and A. Kirrander, Phys. Rev. Lett. {\bf 112}, 043002 (2014)].
\end{abstract}
\maketitle

The main results of Ref.~\cite{SuKi14} rely on the assumption of the validity of Eq.~(1) in Ref.~\cite{SuKi14}. 
In essence, that equation is meant to establish a connection between the time-dependent electron density 
of a nonstationary electronic system and the observable x-ray scattering pattern associated with that system. 
The authors of Ref.~\cite{SuKi14} claim that their Eq.~(1) rests exclusively on the assumption that the 
electronic dynamics to be imaged are much slower than the duration of the x-ray pulse employed to probe 
those dynamics. (This is in addition to the assumption of nonresonant x-ray probe conditions.) The purpose 
of this Comment is to point out that Eq.~(1) in Ref.~\cite{SuKi14} is generally invalid; a short pulse duration 
is by no means sufficient to guarantee the validity of that equation. An example demonstrating the failure of 
Eq.~(1) in Ref.~\cite{SuKi14} for an x-ray pulse shorter than the electron dynamics to be imaged may be found 
in Ref. \cite{DiVe12}. Therefore, we consider the validity of the results of Ref.~\cite{SuKi14} questionable. 
Particularly, it must be expected that the patterns in Fig.~5 of Ref.~\cite{SuKi14} differ qualitatively from 
what would be found in experiment.

Let us assume that the electronic wave packet of interest evolves freely, i.e.,
\begin{equation}
\label{eq3}
|\Psi,t\rangle = \sum_I \alpha_I e^{-i E_I t}| \Psi_I \rangle,
\end{equation}
where $| \Psi_I \rangle $ is an eigenstate of the electronic Hamiltonian, $E_I$ is the associated eigenenergy,
and $\alpha_I$ is a time-independent, generally complex expansion coefficient.
In order to make things particularly transparent, let us assume that only two electronic 
eigenstates---$| \Psi_1 \rangle $ and $| \Psi_2 \rangle $---contribute to the wave packet. Then, the expectation value
of any observable will oscillate periodically with the period $T = 2\pi/|E_2 - E_1|$. A probe pulse that can resolve
these oscillations will necessarily be shorter than $T$; equally necessarily, such a pulse has a spectral
bandwidth that exceeds the energy splitting between $| \Psi_1 \rangle $ and $| \Psi_2 \rangle $. This is 
obvious from Fourier considerations, and it holds irrespective of whether $T$ is a femtosecond or much longer.
It is fundamentally impossible to make the spectral bandwidth of the incoming x-ray beam small in comparison 
to the energy splitting between $| \Psi_1 \rangle $ and $| \Psi_2 \rangle $ if one is using an x-ray pulse that is short 
enough to resolve the dynamics associated with coherent superpositions of those electronic eigenstates. It must therefore
be expected that even if the x-ray scattering detector has perfect energy resolution (a most optimistic assumption), 
the scattering signal will involve an incoherent sum over all final states that are energetically accessible within 
the spectral bandwidth of the incoming x-ray beam. A careful analysis within the framework of quantum electrodynamics 
demonstrates that this is indeed the case \cite{DiVe12}, leading, in general, to a failure of Eq.~(1) in Ref.~\cite{SuKi14}. 

This failure is particularly easy to see if we assume, for simplicity, that the only final states energetically accessible
are $| \Psi_1 \rangle $ and $| \Psi_2 \rangle $ themselves (a gross oversimplification in view of the nonuniform
energy level structure in the Coulomb problem). Then, for a probe pulse much shorter than $T$, the differential 
scattering probability per x-ray pulse, at high photon energy, is 
\begin{equation}
\label{eq4}
\frac{dP}{d\Omega} = \zeta \int d^3x \int d^3x'
\langle \Psi,t_d | \hat{n}({\bm x}) 
\left\{|\Psi_1 \rangle \langle \Psi_1| + |\Psi_2 \rangle \langle \Psi_2|\right\} 
\hat{n}({\bm x'}) | \Psi, t_d \rangle
e^{i{\bm Q}\cdot({\bm x}-{\bm x'})}.
\end{equation}
Here, $\hat{n}({\bm x})$ is the electron density operator, ${\bm Q}$ is the photon
momentum transfer, $t_d$ is the time at which the x-ray probe pulse scatters from the electronic wave packet 
(the pump-probe time delay), and the constant $\zeta$ depends, among other things, on the spectrum of the incoming 
x-ray beam and on the spectral 
response of the x-ray scattering detector. Equation~(\ref{eq4}) may be easily verified by using the 
results of Ref.~\cite{DiVe12}. (One may arrive at the same conclusion by applying the analyses of Refs.~\cite{TaCh01} and 
\cite{HeMo08}.) The key point here is that the right-hand side of Eq.~(\ref{eq4}) {\em cannot} be written in the form
\[
\int d^3x \int d^3x'                                    
\langle \Psi,t_d | \hat{n}({\bm x}) | \Psi, t_d \rangle \langle \Psi,t_d | 
\hat{n}({\bm x'}) | \Psi, t_d \rangle
e^{i{\bm Q}\cdot({\bm x}-{\bm x'})},
\]
which, up to a prefactor, is Eq.~(1) from Ref.~\cite{SuKi14} in the notation employed here. In other words, the requirement 
of a short pulse---or slow electronic dynamics---does not ensure that the final state reached in the photon 
collision process equals the electronic wave-packet state right before the collision. Finally, we would like to mention
that analogous considerations have been shown to apply to time-resolved electron scattering \cite{ShSt13}.

\end{document}